\documentclass{ws-ijmpa}

\begin{document}


%
%

\title{ Role of Higher Fermion Representations in TeV-scale Seesaw\footnote{Talk at BUE-CTP International Conference on Neutrino Physics in the LHC Era, Luxor, Egipt, November 2009.}
}

\author{Ivica~Picek\footnote{picek@phy.hr} ~ and  Branimir~Radov\v{c}i\'c\footnote{bradov@phy.hr}}

\address{Department of Physics, Faculty of Science, University of Zagreb,
 P.O.B. 331\\ HR-10002 Zagreb, Croatia\\}

\maketitle


\begin{abstract}

We consider a scenario in which additional vectorlike TeV-scale fermions belonging to higher weak-isospin multiplets 
provide new seesaw mediators. If these fermions have non-zero hypercharge, their tree-level exchange 
produces novel seesaw mechanism different from type I and III seesaw.
In order to produce Majorana masses for light 
neutrinos, new Dirac seesaw mediators are constrained by the SM gauge symmetry to belong 
to a weak triplet and a five-plet. 
The latter, in conjunction with two isospin 3/2 scalar multiplets,
leads to new seesaw formula $m_{\nu} \sim v^6/M^5$. It reproduces the  empirical 
masses $m_{\nu} \sim 10^{-1}$ eV by $M \leq$ TeV new states, testable at the LHC.

\end{abstract}

\ccode{PACS numbers: 14.60.Pq; 14.60.St; 12.60-i}

\section{Landmarks}

The start-up of the Large Hadron Collider (LHC) opens for exploration the upcoming landscape of physics.
Our hope is that a strangeness standing up in the landscape of the Standard Model (SM)
gives us ability to address the physics beyond the Standard Model (BSM). Three such landmarks  are:
\begin{itemlist}
 \item Lightness of neutrinos,
 \item Heaviness of the top quark,
 \item Lightness of the SM Higgs doublet.
\end{itemlist}
The last landmark with special prominence, known as the {\em hierarchy problem}, reflects an inexplicable
quantum stability of the weak
scale {\em v} with respect to the UV cutoff. In order to tame the quadratically divergent
contributions to the Higgs mass, various BSM approaches have been introduced, like little Higgs, supersymmetry, technicolor and extra dimensions.

The extensions of the SM aimed at explaining lightness of neutrinos came as if on cue for the GUT programme \cite{Georgi:1974sy}. By integrating out the GUT-scale degrees of freedom one arrives at the tree level at
neutrino masses through an effective dimension-five operator\cite{Weinberg:1979sa} $LL\Phi\Phi$. The heavy fermion singlet and triplet mediators result in the dimension-five operator through type I and type III seesaw mechanisms\cite{Minkowski:1977sc-etc,Foot:1988aq}. They produce small neutrino mass  $m_\nu\sim v^2/M$ at the cost of introducing
a new hierarchy problem of the seesaw scale $M \sim 10^{14}$ GeV with respect to $v$. By admitting some of the GUT-scale degrees of freedom to be as low as the TeV scale as in a recent reincarnation of $SU(5)$ GUT model\cite{Bajc:2006ia}, one arrives at a low scale hybrid type I and III seesaw model. Multiple seesaw approach introduced by Ma\cite{Ma:2000cc} and extended\cite{Grimus:2009mm} later, brings the seesaw mechanism to the TeV scale\cite{Xing:2009hx} accessible at the Large Hadron Collider (LHC). Recent lowering\cite{Babu:2009aq} of the seesaw scale by dim = 7 operator and  with a focus on the triple charged scalars has led to $m_{\nu} \sim v^4/M^3$. This study has followed an earlier observation of the efficiency of higher scalar multiplets in suppressing neutrino masses\cite{Tavartkiladze:2001by}.

Here we describe our recent work\cite{Picek:2009is} which aims at a lowering the seesaw scale in a ``bottom-up'' approach, by extending the SM particle content by higher then SM weak isospin multiplets.
 An important assumption in our approach is that new fermionic degrees of freedom beyond the three SM 
generations are realized as vectorlike Dirac states. We examined recently\cite{Picek:2008dd} similar states in the quark sector represented by the  vectorlike top partner which has dispayed the mass matrix  corresponding to a Dirac seesaw mechanism. In the following we will  describe how this model explains the large mass of the top quark\cite{Vysotsky:2006fx}, and compare it with analogous Dirac seesaw mechanism appearing in the particular extension\cite{Ma:2002tc}
of the supersymmetric SM. After that we  present our novel seesaw mechanism\cite{Picek:2009is} for generating Majorana masses of light neutrinos by Dirac mediators, which is different from type I and III seesaw mechanisms where both light neutrinos and heavy messengers are represented by Majorana particles.

\subsection{Dirac seesaw with terascale vectorlike top-partner}

We elaborated recently\cite{Picek:2008dd} on the model with a vectorlike top-partner
 $T$ proposed by Vysotsky\cite{Vysotsky:2006fx}. Expressed in terms of the weak (primed)
eigenstates, it reads in the form of the Lagrangian that in addition to
the usual SM piece has the BSM part consisting of two Dirac
mass terms and one Yukawa term:
\begin{equation}
{\cal L}_{BSM} = -M \bar T^\prime_L T^\prime_R
+ \left[
\mu_R \bar T^\prime_L t^\prime_R
+ \frac{\mu_L}{v/\sqrt 2} \left(\bar t^\prime, \bar b^\prime \right)_L \Phi^c \; T_R^\prime \right] + h.c. \;\; .
\label{1}
\end{equation}
Two new heavy $SU(2)_L$ singlet states, $T^\prime_{L}$ and
$T^\prime_{R}$, have the Dirac mass term $M$. The $T^\prime - t^\prime$ mixing is
given by two terms in the square brackets: the $\mu_R$ term
describing the mixing of two $SU(2)_L$ singlets, $T_L^\prime$ and
$t_R^\prime$, and the $\mu_L$ term describing mixing of the SM weak
isodoublet with the isosinglet state $T^\prime$. Obviously, by
switching off these $\mu_{L,R}$ terms,
the $t^\prime$ field would become the ordinary $t$ quark, the mass
eigenstate of the SM.

The presence of the $T^\prime-t^\prime$ mixing has several effects examined before\cite{Picek:2008dd,Vysotsky:2006fx}. Here we stress that, in order to find the states with definite masses the following matrix should be diagonalized:
\begin{equation}
M_{t-T}=(\overline{t_L^\prime} \overline{t_R^\prime} \overline{T_L^\prime}
\overline{T_R^\prime}) \left( \begin{array}{cccc} 0 & m_t & 0 &
\mu_L \\ m_t & 0 & \mu_R & 0 \\ 0 & \mu_R & 0 & -M \\ \mu_L & 0 &
-M & 0 \end{array} \right) \left( \begin{array}{l} t_L^\prime \\
t_R^\prime \\ T_L^\prime \\ T_R^\prime \end{array} \right) \;\; .
\label{mass_matrix_top}
\end{equation}
From here we can explain the heaviness of top\cite{Vysotsky:2006fx}. By switching of 
the mass of $t$ quark in SM, i.e. by putting $m_t =0$, top quark massless in SM gets all
its mass $\sim \mu_L \mu_R / M$ from the mixing with heavy $T$.

\subsection{Dirac seesaw for neutrinos in model by E. Ma}

The seesaw mechanism resembling the one based on eq.(\ref{mass_matrix_top}) appears in neutrino sector in the particular extension
of the supersymmetric SM proposed by Ma\cite{Ma:2002tc}. He added a new $U(1)_X$ gauge symmetry and new supermultiplets at the TeV scale to solve both the $\mu$ problem and the absence of the neutrino masses in the MSSM. Additional isospin singlet fermions mix with light neutrinos and span the following mass matrix in the basis $(\nu,S^C,N,N^C)$
\begin{equation}
M_\nu =  \left( \begin{array}{cccc} 0 & 0 & 0 &
m_1 \\ 0 & 0 & m_2 & 0 \\ 0 & m_2 & 0 & M \\ m_1 & 0 &
M & 0 \end{array} \right) \;\; ,
\label{mass_matrix_ma}
\end{equation}
where $N$ and $N^C$ are independent fields.
In this model the lepton number is conserved and the light neutrinos are Dirac fermions with mass given by $m_1 m_2 / M$. 
Here the mass $m_1$ comes from the weak scale {\em v},  $m_2$ comes from $U(1)_X$ symmetry breaking, and $M$ is an invariant mass. Because both the light neutrinos and the heavy neutral states are Dirac fermions, this mechanism is called a Dirac seesaw. 

After the examples above, we can explore the possibility of a viable seesaw model with  light Majorana neutrinos by employing heavy Dirac neutral states.

\section{Model for light Majorana neutrinos with  TeV-scale
Dirac mediators}

In order to have Dirac seesaw mediators we have to departure from hypercharge zero multiplets used in type I and III seesaw mechanism. This   can be achieved by invoking  vectorlike fermionic multiplets which, although not restricted to the electroweak scale, could be light enough to avoid the hierarchy problem. In order to have a tree level seesaw, the newly introduced fermion multiplets have to contain a neutral component which mixes with light neutrinos. Additional scalar multiplets have to be introduced in order to have Yukawa terms necessary for fermionic tree level seesaw. This leads to restricted quantum number assignments for new fermionic multiplets: the  vectorlike triplet\cite{Babu:2009aq} and the vectorlike five-plet\cite{Picek:2009is} corresponding to effective dim = 7 and dim = 9 operators, which are additional to those presented in a previous study\cite{Bonnet:2009ej}.

\subsection{Fermion - Yukawa Sector}

Our model\cite{Picek:2009is} obeys the SM gauge group $SU(3)_C\times SU(2)_L \times U(1)_Y$ symmetry. The fermions additional to those of the SM belong to vectorlike Dirac five-plet $\Sigma$ of leptons, so that both their left $\Sigma_L = (\Sigma^{+++}_L, \Sigma^{++}_L, \Sigma^+_L, \Sigma^0_L, \Sigma^-_L)$ and right $\Sigma_R = (\Sigma^{+++}_R, \Sigma^{++}_R, \Sigma^+_R, \Sigma^0_R, \Sigma^-_R)$ components transform as $(1,5,2)$. The vectorlike $\Sigma$  can form a gauge invariant Dirac mass term
\begin{equation}
\mathcal{L}_\text{mass} =  -\,  M_{\Sigma} \, \overline \Sigma_L \Sigma_R +   \text{H.c.} \ .
\label{Dir}
\end{equation}
The scalar multiplets needed for gauge invariant Yukawa terms,
\begin{equation}
\mathcal{L}_{\text{Y}} = Y_1 \, \overline l_{L} \Sigma_R
\, \Phi_1  +Y_2 \, \overline \Sigma_L (l_{L})^c \, \Phi^*_2   +\text{H.c.} \ ,
\label{Yuk}
\end{equation}
are weak isospin $3/2$ scalar 4-plets $\Phi_1=(\Phi^{0}_1,
\Phi^{-}_1, \Phi^{--}_1, \Phi^{---}_1)$ and $\Phi_2=(\Phi^{+}_2,
\Phi^{0}_2, \Phi^{-}_2, \Phi^{--}_2)$ transforming as $(1,4,-3)$ and $(1,4,-1)$, respectively.
The vacuum expectation values ({\em vevs}) of $\Phi_1$ and $\Phi_2$ fields in the Yukawa terms in eq.~(\ref{Yuk}) result in 
the mass terms which connect the lepton doublet with new vectorlike lepton 5-plet.

If we restrict ourself to one SM lepton doublet $l_L$, the three neutral left-handed components $\nu_L$, $\Sigma_L^{0}$ and $(\Sigma_R^{0})^c$ span the mass matrix
\begin{eqnarray}
\mathcal{L}_{\nu \Sigma^0} & = &  \, -\frac{1}{2}
\left(\bar \nu_L \; \overline{\Sigma_L^0} \; \overline{(\Sigma_R^0)^c} \right)
\left( \! \begin{array}{ccc}
0 & m_2 & m_1 \\
m_2 & 0 & M_{\Sigma} \\
m_1 & M_{\Sigma} & 0
\end{array} \! \right) \,
\left( \!\! \begin{array}{c} (\nu_L)^c \\ (\Sigma_L^0)^c \\ \Sigma_R^0 \end{array} \!\! \right)
\; + \mathrm{H.c.}\ .
\label{neutral_mass_matrix}
\end{eqnarray}
Here $m_1$ and $m_2$ result from the first and second term of eq.~(\ref{Yuk}), respectively, and $M_{\Sigma}$ is given by eq.~(\ref{Dir}). Consequently, $m_1$ and $m_2$ are on the scale of the {\em vev} $v_{\Phi_1}$ and {\em vev} $v_{\Phi_2}$ of the neutral components of the scalar quadruplets, but $M_{\Sigma}$ is on the new physics scale $\Lambda_{NP}$ larger then the electroweak scale.
After diagonalizing the mass matrix in eq.~(\ref{neutral_mass_matrix}), the  light neutrino acquires a Majorana mass
\begin{equation}
m_{\nu} \sim \frac{m_1m_2}{M_\Sigma}\ .
\label{seesaw}
\end{equation}
Starting with heavy Dirac lepton states, their mixing with  both $l_L$ and $(l_L)^C$ results in light massive Majorana  neutrino.
Therefore, the lepton number is violated in our model. The difference between the type I and III seesaw and our seesaw model can be understood by looking at the lepton number violation (LNV). Despite a freedom in assigning lepton numbers to new multiplets, our vectorlike non-zero hypercharge seesaw messengers do not allow a lepton number assignment that would bring the LNV to the mass term as is the case in type I and III seesaw.

\subsection{Scalar Sector}

Turning to the scalar potential, we restrict ourselves to the following renormalizable terms relevant for our mechanism:
\begin{eqnarray}\label{pot}
V(H, \Phi_1, \Phi_2) &\sim& \mu_H^2 H^\dagger H + \mu^2_{\Phi_1} \Phi^\dagger_1 \Phi_1+ \mu^2_{\Phi_2} \Phi^\dagger_2 \Phi_2 + \lambda_H (H^\dagger H )^2 \nonumber \\
 &+& \{ \lambda_1 \Phi^*_1 H^* H^* H^* + \mathrm{H.c.} \} + \{ \lambda_2 \Phi^*_2 H H^* H^* + \mathrm{H.c.} \} \nonumber\\
 &+& \{ \lambda_3 \Phi^*_1 \Phi_2 H^* H^* + \mathrm{H.c.} \}\ .
\end{eqnarray}
For the {\em vevs} of the doublet and two 4-plets we assume the values:
\begin{equation}\label{vev}
    \langle 0|H|0\rangle = \left( \begin{array}{c} 0 \\ v \end{array} \right)\ ,\ \langle 0|\Phi_1|0\rangle 
= \left( \begin{array}{c} v_{\Phi_1} \\ 0 \\ 0 \\ 0 \end{array} \right)\ ,\ \langle 0|\Phi_2|0\rangle = \left( \begin{array}{c} 0 \\ v_{\Phi_2} \\ 0 \\ 0 \end{array} \right)\ .
\end{equation}
Inserting these {\em vevs} into the vacuum value $\langle 0|V(H, \Phi_1, \Phi_2)|0\rangle = V_0 $, we obtain
\begin{eqnarray}\label{potvev}
V_0 &\sim& \mu_H^2 v^2 + \mu^2_{\Phi_1} v_{\Phi_1}^2 + \mu^2_{\Phi_2} v_{\Phi_2}^2 + \lambda_H v^4 \nonumber \\
 &+& \{ \lambda_1 \sqrt{\frac{1}{4}} v_{\Phi_1}^* v^{*3} + \mathrm{H.c.} \} + 
\{ \lambda_2 \sqrt{\frac{1}{6}} v_{\Phi_2}^* v^{*3} + \mathrm{H.c.} \} + {\cal O} (v^2_{\Phi_i})\ ,
\end{eqnarray}
where the numerical factors $\sqrt{\frac{1}{4}}$ and $\sqrt{\frac{1}{6}}$ are Clebsch-Gordan coefficients.

The conditions for the minima, $\partial V_0 / v=0$ and $\partial V_0 / v_{\Phi_i}=0$, give the equations for the vacuum stability:
\begin{eqnarray}
  0 &=& 2 \mu_H^2 v + 4 \lambda_H v^3 + {\cal O} (v_{\Phi_i})\ ,\\
  0 &=& \mu^2_{\Phi_1} v_{\Phi_1} + \lambda_1 \sqrt{\frac{1}{4}} v^3 + {\cal O} (v^2_{\Phi_i})\ , \\
  0 &=& \mu^2_{\Phi_2} v_{\Phi_2} + \lambda_2 \sqrt{\frac{1}{6}} v^3 + {\cal O} (v^2_{\Phi_i})\ .
\end{eqnarray}
The electroweak symmetry breaking proceeds in the usual way from the {\em vev} $v$ of the Higgs doublet, implying $\mu_H^2<0$ and
\begin{equation}
    v=\sqrt{\frac{-\mu_H^2}{2\lambda_H}}\ .
\end{equation}
Despite the positiveness of $\mu^2_{\Phi_1}$ and $\mu^2_{\Phi_2}$, there are non-vanishing induced {\em vevs} $v_{\Phi_1}$ and $v_{\Phi_2}$, approximately given by
\begin{equation}\label{phivev}
    v_{\Phi_1}=-\lambda_1 \sqrt{\frac{1}{4}} \frac{v^3}{\mu^2_{\Phi_1}}\ ,\ v_{\Phi_2}=
-\lambda_2 \sqrt{\frac{1}{6}} \frac{v^3}{\mu^2_{\Phi_1}}\ .
\end{equation}
By merging eq.~(\ref{Yuk}), eq.~(\ref{seesaw}) and eq.~(\ref{phivev}) we obtain the light neutrino mass
\begin{equation}\label{dim9}
m_{\nu}  \sim \frac {Y_1 Y_2 v_{\Phi_1} v_{\Phi_2}} {M_{\Sigma}} \sim \frac {Y_1 Y_2\ \lambda_1\lambda_2\ v^6} {M_{\Sigma}\ \mu^2_{\Phi_1}\ \mu^2_{\Phi_2}} \,.
\end{equation}
The {\em vevs} of our scalar multiplets change the $\rho$ parameter to $\rho (\Phi_1) \simeq 1-6v^2_{\Phi_1}/v^2$ and $\rho (\Phi_2) \simeq 1+6v^2_{\Phi_2}/v^2$. By matching $\rho (\Phi_1)$ and $\rho (\Phi_2)$ to the experimental value\cite{PDG08} $\rho=1.0000^{+0.0011}_{-0.0007}$ , we get the upper bounds $v_{\Phi_1}\leq1.9$ GeV and $v_{\Phi_2}\leq2.4$ GeV.

\section{Testability}

We can estimate the high energy scale of our model from eq.~(\ref{dim9}) by assuming the same value for the mass parameters, $\mu_{\Phi_1} \simeq \mu_{\Phi_2} \simeq M_\Sigma \simeq \Lambda_{NP}$. By taking values $v=174$ GeV and $m_\nu\sim0.1$ eV we obtain 
$\Lambda_{NP}\simeq580$ GeV for $Y_1\sim Y_2\sim \lambda_1\sim \lambda_2\sim10^{-2}$.

We also have two types of loop contributions, one steaming from closing $(H^\dagger H)$ legs in the tree-level diagram,
and the other generated by $\lambda_3$ term in eq. (\ref{pot}). With characteristic loop suppression factor, the first loop contribution is smaller then the tree level one if\cite{Gouvea:2008} $\Lambda_{NP}<4 \pi v\simeq2$ TeV, and the second if $\Lambda_{NP}  <\sqrt{4 \pi} v \simeq 620$ GeV. Obviously, the later numerical value has to be taken conditionally, and has a meaning of a bound of a few 100 GeV, meaning that all new states in our model should lie below $\sim$ TeV. Such states are expected to be abundantly produced at hadronic colliders through $W$, $Z$ and $\gamma$ Drell-Yan fusion processes. 

The associated production of the pairs $(\Sigma^{+++},\overline{\Sigma^{++}})$, $(\Sigma^{++},\overline{\Sigma^{+}})$, $(\Sigma^{+},\overline{\Sigma^{0}})$, $(\Sigma^{0},\overline{\Sigma^{-}})$ via a charged current is a crucial test of the five-plet nature of new leptons. Direct pair production of neutral states $(\Sigma^{0},\overline{\Sigma^{0}})$ via a neutral current, which is not possible for type I and III mediators, is possible for our states with non-zero weak charges.

Vectorlike fermions of the type considered here are characterized by small mass splitting within a multiplet\cite{Cirelli:2005uq}. The hierarchy within the $\Sigma$ multiplet can be read out from\cite{Cirelli:2005uq} 
\begin{equation}\label{mass_diffrence}
    M_Q-M_0\simeq Q(Q+Y/\cos\theta_W)\Delta M\ ,
\end{equation}
where $\Delta M$ is the mass splitting between $Q=1$ and $Q=0$ components in the case of zero hypercharge:
\begin{equation}\label{mass_delta}
    \Delta M = \alpha_2 M_W \sin^2\frac{\theta_W}{2}=(166\pm1) MeV.
\end{equation}
We find that $\Sigma^-$ is the lightest component of the 5-plet, being 210 MeV lighter then $\Sigma^0$. Additional small mass splitting\cite{Grimus:2000vj} within the  5-plet  comes from the mixing of the 5-plet components with the SM leptons, but this is of the order of $m_1 m_2/ M_\Sigma \sim 0.1$ eV which is totally negligible.

Let us finally list the decays providing  distinctive signatures of $\Sigma$ states at the colliders. A triply charged $\Sigma^{+++}$ decays through an off-shell $\Sigma^{++}$, leading typically to $\Sigma^{+++} \to W^+ W^+ l^+$ decay. A doubly charged state decays as $\Sigma^{++} \to W^+ l^+$ The singly charged states decays as $\Sigma^{+} \to W^+ \nu,Z l^+,H^0 l^+$ and $\Sigma^{-} \to W^- \nu,Z l^-,H^0 l^-$. Finally, a neutral seesaw mediator decays as $\Sigma^{0} \to W^\pm l^\mp, Z \nu, H^0 \nu$.

To conclude, there is a number of processes\cite{Picek:2009is}
which should point to the existence of the 5-plet states under consideration, if they explain
the neutrino masses at the tree level. It is gratifying that the parameter space corresponding to the novel tree-level model will be tested at the LHC, so that this model is falsifiable. A very non-discovery of the fermionic 5-plet states would leave us with a possibility of their role in neutrino mass generation at the loop-level.

\section*{Acknowledgments}

We thank Goran Senjanovi\'c for critical remarks and B.R. thanks Borut Bajc for useful discussions. Both of us thank Walter Grimus for hospitality offered at the University of Vienna. This work is supported by the Croatian Ministry  of Science, Education and Sports under Contract No. 119-0982930-1016.


\end{document}